\documentclass[]{spie}  

\usepackage{amsmath,amsfonts,amssymb}
\usepackage{booktabs}
\usepackage{float}
\usepackage{graphicx}
\usepackage{gensymb}
\usepackage[colorlinks=true, allcolors=blue]{hyperref}
\usepackage{subcaption}

\usepackage{lmodern}
\usepackage{siunitx}
\usepackage{comment}

\title{Focal Plate Prototyping for Modular Focal Planes of Stage-5 Instruments For Ground-Based Telescopes}

\author[a]{Maxime Rombach}
\author[b]{Jean-David Perriard}
\author[c]{Laurent Chevalley}
\author[a]{Diane Chapuis}
\author[d]{Markus Thurneysen}
\author[a]{Jean-Paul Kneib}
\affil[a]{Institute of Physics, Laboratory of Astrophysics, Ecole Polytechnique Federale de Lausanne (EPFL), Observatoire de Sauverny, CH-1290 Versoix, Switzerland}
\affil[b]{Institute of Physics Workshop, Ecole Polytechnique Federale de Lausanne (EPFL), BSP 210, CH-1015 Lausanne}
\affil[c]{Institute of Mechanical Engineering Workshop (ATME), Ecole Polytechnique Federale de Lausanne (EPFL), MEC 1406, CH-1015 Lausanne}
\affil[d]{Haute Ecole du Paysage, d'Ingénierie et d'Architecture de Genève (HEPIA), Hautes Ecoles Spécialisées de Suisse Orientale (HES-SO), Rue de la prairie 4, 1202 Geneva, Switzerland}

\authorinfo{Further author information: (Send correspondence to M. R.)\\M.R.: E-mail: maxime.rombach@epfl.ch, Telephone: +41 (0) 21 693 28 76}

\begin{document} 
\maketitle
\thispagestyle{empty}

\begin{abstract}
As current Stage-4 multi-object instruments such as SDSS-V, DESI, MOONS or 4MOST are providing astrophysicists data to study the objects of the Universe, effort is arising to build the next generation of Stage-5 multi-object focal planes; aiming for 20’000 fibers-class instruments. The focal plate structure is a central element of the future focal plane assemblies. It maintains the fiber positioners, the Guide, Focus and Alignment cameras (GFAs) and wave-front sensors together on the focal surface of the telescope. In addition to being optimized for stiffness and mass, the plate needs to meet tight tolerances in tilt, typically ± 0.05 $\degree$, and focus, typically, ± 30 $\mu$m, to match the telescope’s curved focal surface.
The presented focal plate prototype shows that 5-axes machining is promising to meet the desired tolerances.

\end{abstract}


\section{INTRODUCTION}
\label{sec:intro}  

The next generation of Stage-5 astronomical surveys aims for a higher fiber positioners multiplex than ever before. About 20’000 fiber positioners will sit in the focal plane of the telescopes depending on the project (MUST\cite{cai_large_2025}, Spec-S5\cite{besuner_spectroscopic_2025}, or WST\cite{mainieri_wide-field_2024}). The fiber positioners assembly will likely be \textit{modularized} i.e. the robots will be grouped into groups in standalone sub-assemblies. EPFL is currently developping a \textit{triangular module} approach that packages the positioners in groups of 63 separated by a \textit{pitch} distance of 6.2mm\cite{galal_prototyping_2026}.  First theorized in Silber, et al. (2022) \cite{silber_25000_2022}, further research to optimize the modules layout has been done in Rombach, et al. (2024) \cite{rombach_investigations_2024}. The latter concluded that 63 positioners per module was an optimum solution. The current study aims to follow-up this work and shares the steps towards manufacturing a \textit{focal plate structure} to secure the modules in the instrument.

\begin{figure}[H]
     \centering
     \begin{subfigure}[b]{0.49\textwidth}
        \centering
        \includegraphics[width=0.6\linewidth]{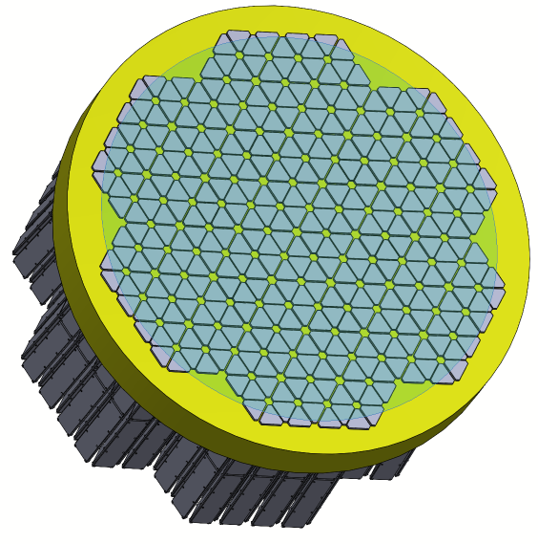}
        \caption{Focal plane modular architecture concept ($\approx$ 180 to 460 modules depending on the instrument project)}
        \label{fig:MM_FP}
     \end{subfigure}
     \hfill
     \begin{subfigure}[b]{0.49\textwidth}
         \centering
         \includegraphics[width=0.9\textwidth]{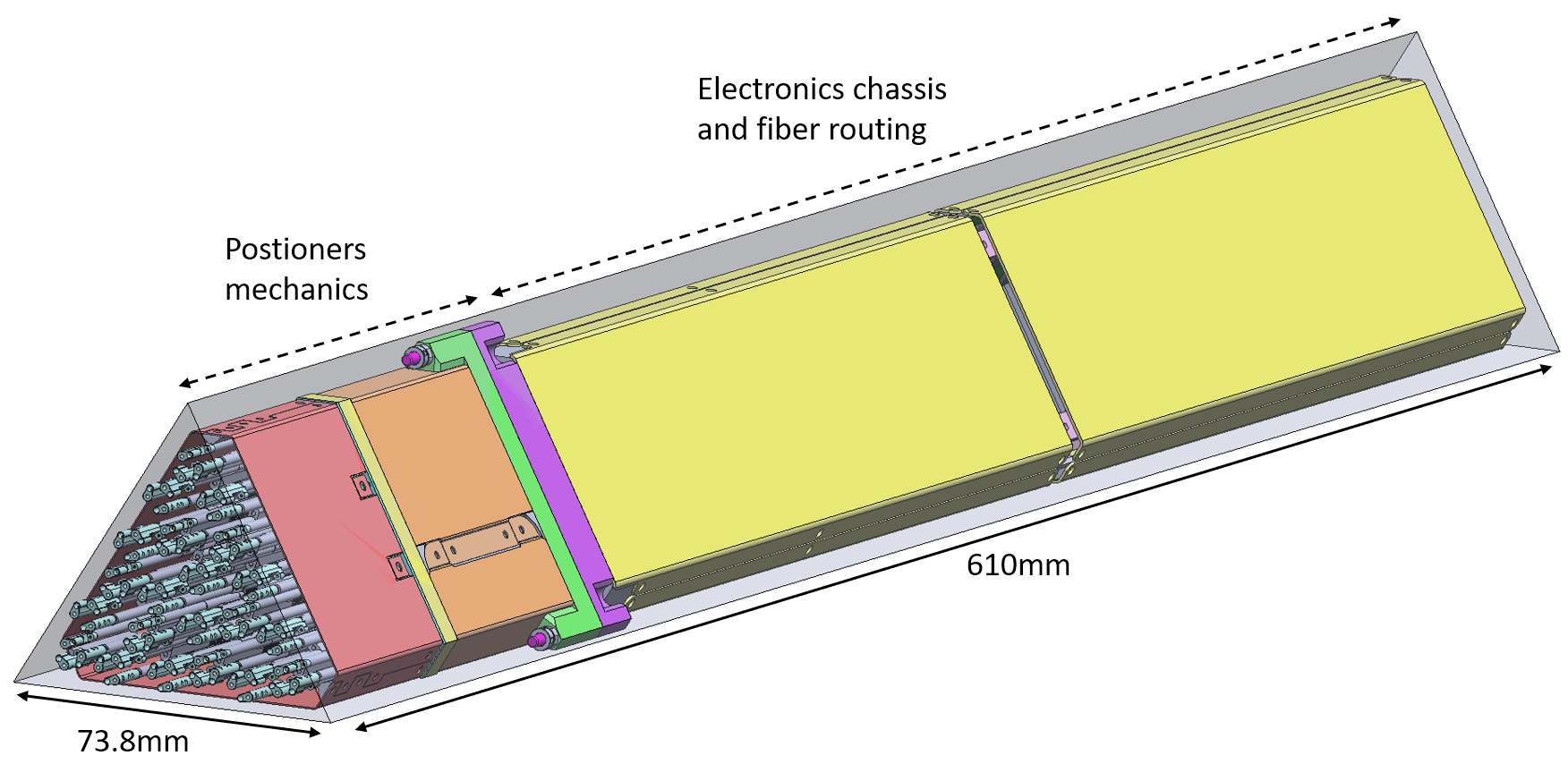}
         \caption{Triangular module CAD model overview; the module fits in an prismatic envelope of 73.8 x 610 mm}
         \label{fig:module_envelope}
     \end{subfigure}
     \vspace{0.2cm}
        \caption{Modular focal plane concept and triangular module concept}
        \label{fig:Intro_images}
\end{figure}
\section{Plate design \& Manufacturing}
\subsection{Focal surface parameters}
Since this work was done in the scope of the \textit{Innosuise project} -  No. 101.014 IP-ENG, it was not tied to any of the current future Stage-5 instrument project. Therefore, design choices were made in order to simplify the prototyping process whilst staying as close as possible to a real-use case. For instance we wanted to optimize for cost and manufacturing time. Hence, we chose to prototype a plate with dimensions matching the Spec-S5 focal surface described in Besuner, et al. (2025)\cite{besuner_spectroscopic_2025}, since those dimensions were matching the largest available CNC machine available in EPFL. Table \ref{tab:focal_surf_param} summarizes those base dimensions along with the few simplification made; namely the focal surface is considered spherical and Chief Ray Deviation (CRD) is neglected, although we aknowledge its significant impact on the nominal orientaiton of the modules. Indeed this work aims to first test the manufacturing feasibility and reachable precision for the modules support faces, which are the cornerstone of the module attachment principle, described in Sec. \ref{sec:focal_plate_mod}.

\begin{table}[H]
\centering
\caption{Focal surface parameters used for this prototype}
\label{tab:focal_surf_param}
\begin{tabular}{llc} 
\toprule
 & \textbf{Value} & \textbf{Unit} \\
Diameter & 818.4 & mm \\
Curvature radius & 13000 & mm \\
Shape & Spherical - Concave & - \\
Chief Ray Deviation & Assumed none & - \\
\bottomrule
\end{tabular}
\end{table}
\subsection{Focal plate modeling}
\label{sec:focal_plate_mod}
\begin{figure}[H]
\captionsetup[subfigure]{justification=centering}
 \begin{subfigure}[t]{0.5\textwidth}
		\centering
		\includegraphics[width=0.9\linewidth]{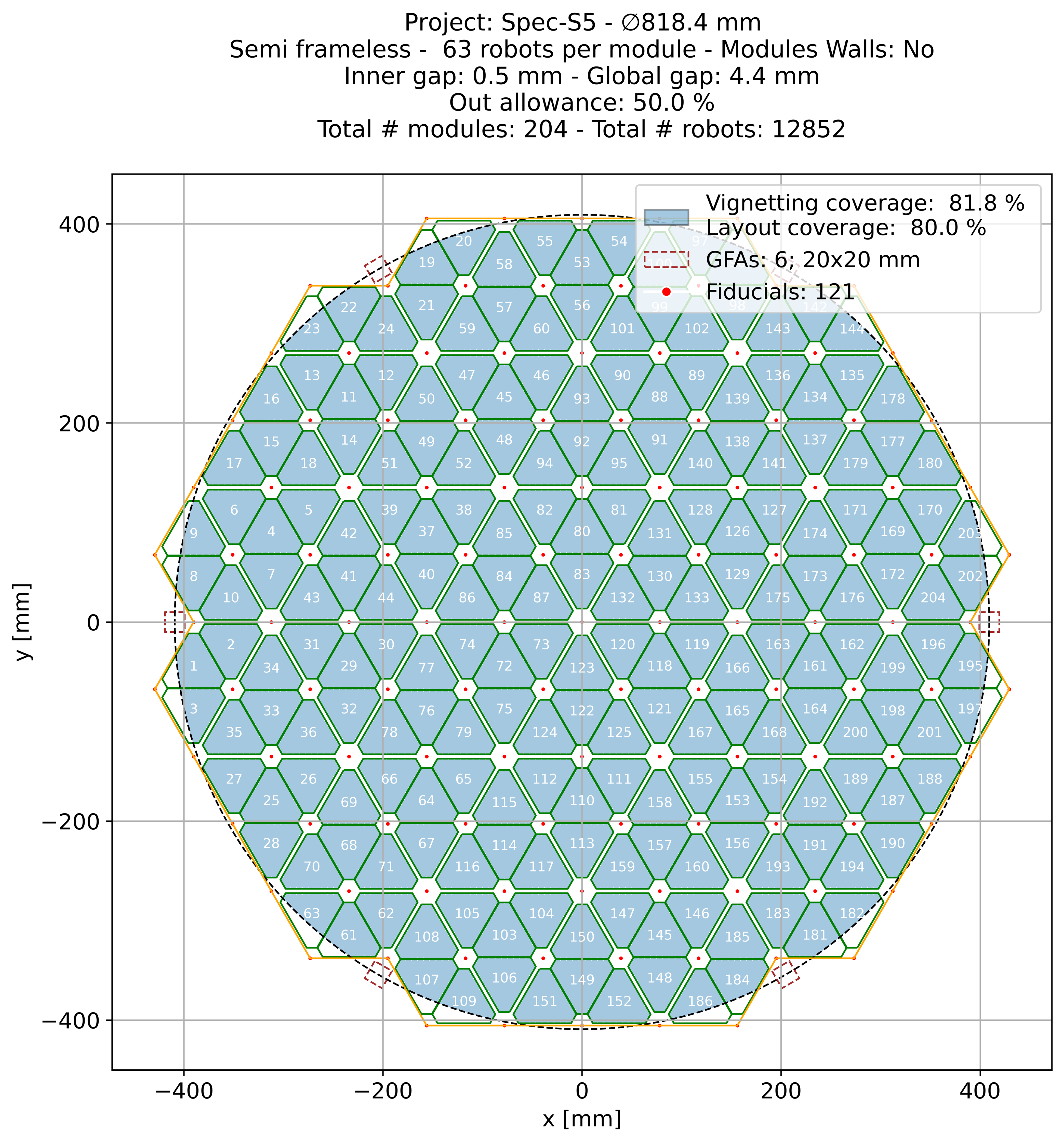}
		\caption{Fully populated focal plane}
        \label{fig:002_FP_layout_360deg}
	\end{subfigure}
	\begin{subfigure}[t]{0.5\textwidth}
		\centering
		\includegraphics[width=0.9\linewidth]{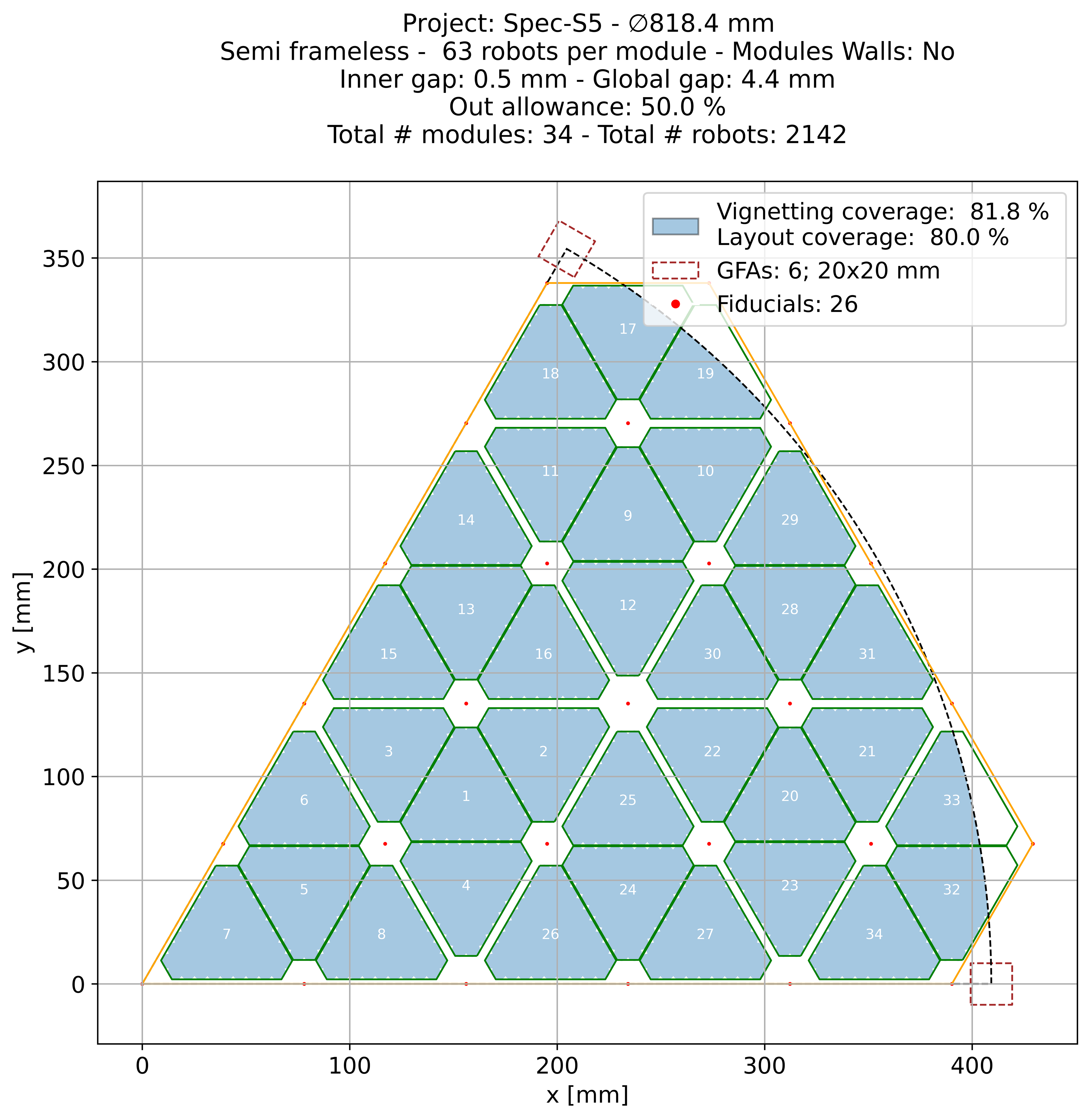}
        \caption{60° slice of the layout; 34 modules}
        \label{fig:002_FP_layout_60deg}
	\end{subfigure}
\vspace{0.2cm}
 \caption{Possible focal plane layout with the given parameters}
 \label{fig:002_FP_layout}
\end{figure}

Modeling this prototype was done with the same automated workflow described in Rombach, et al. (2024)\cite{rombach_investigations_2024}. A custom python script inspired from Silber, et al. (2022)\cite{silber_25000_2022} takes-on the focal surface parameters from Table \ref{tab:focal_surf_param} and the module dimensions to provide a focal plane layout. Figure \ref{fig:002_FP_layout_360deg} shows a possible module layout that optimizes for coverage within the focal surface diameter and boldly assumes sky sensors such guiding cameras and wavefront sensors to be small 20x20mm placeholders.\\
Figure \ref{fig:002_FP_layout_60deg} isolates a 60° slice of the full layout which we will use as a base to build the focal plate prototype.

\begin{figure}[H]
\captionsetup[subfigure]{justification=centering}
 \begin{subfigure}[t]{0.5\textwidth}
		\centering
		\includegraphics[width=0.9\linewidth]{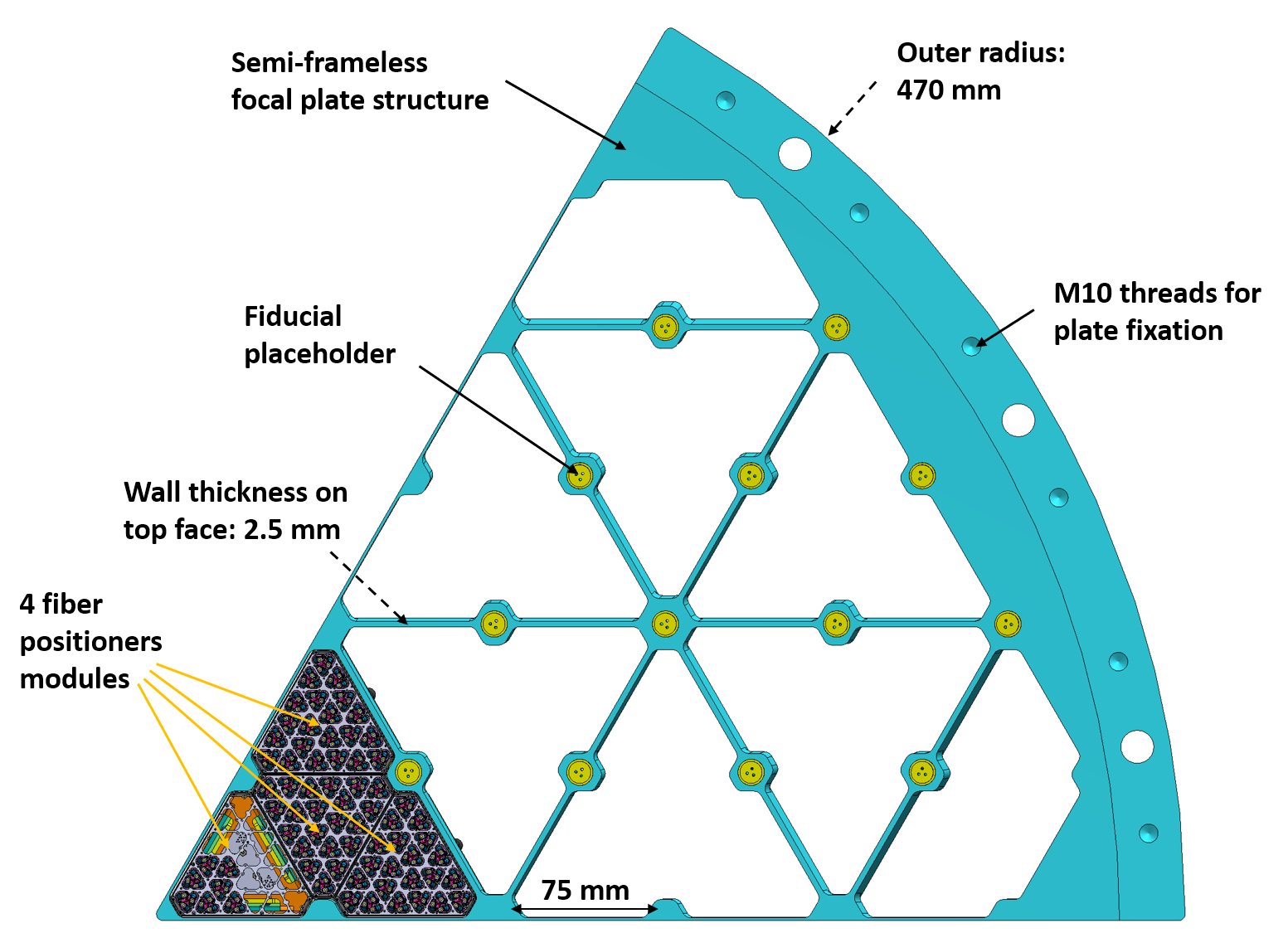}
		\caption{CAD model top view: overview of the semi-frameless focal plate structure prototype with 4 modules assembled}
        \label{fig:002_plat_ass_overview}
	\end{subfigure}
	\begin{subfigure}[t]{0.5\textwidth}
		\centering
		\includegraphics[width=0.9\linewidth]{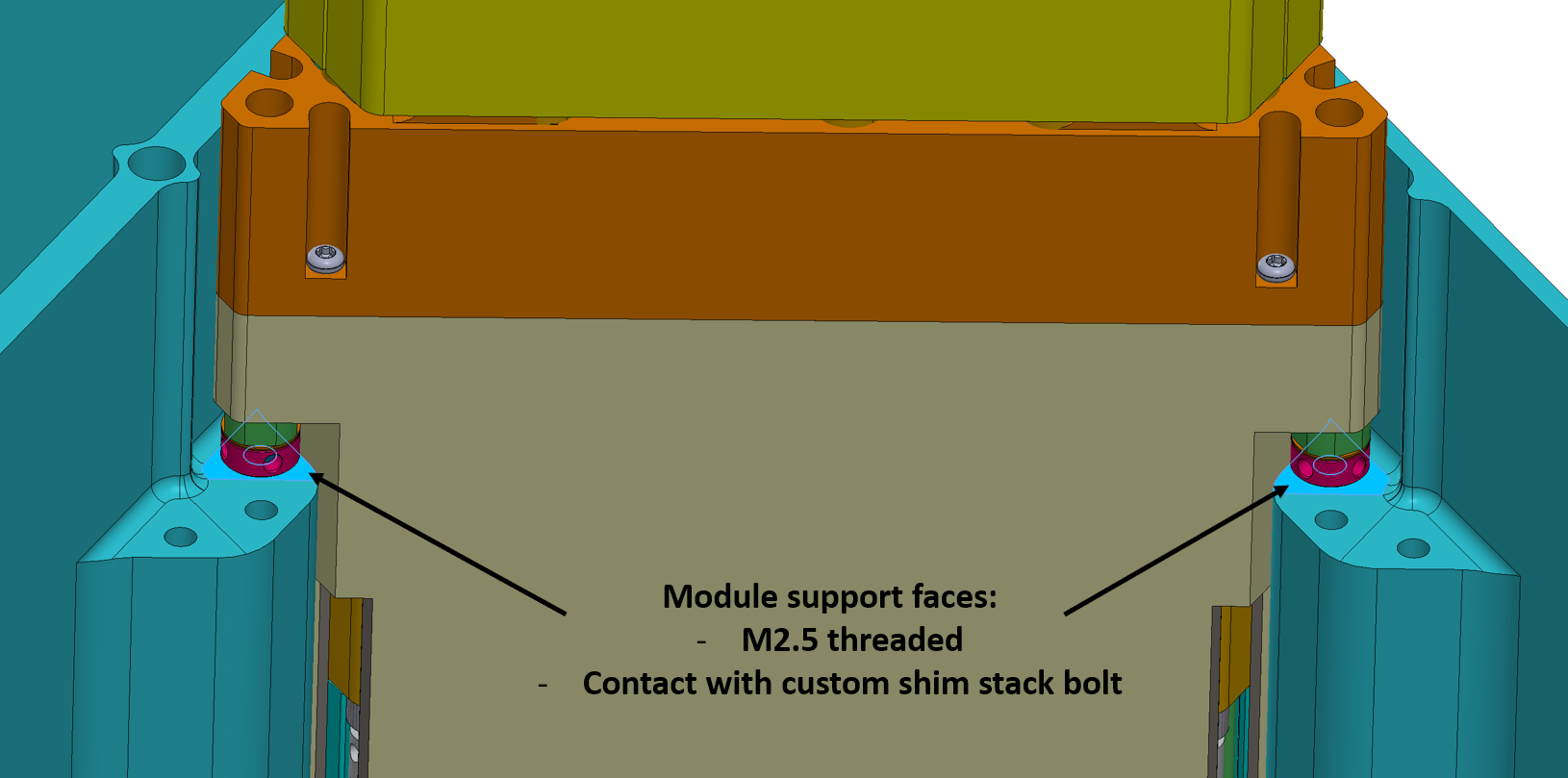}
        \caption{CAD model bottom view: zoom on the module support faces}
        \label{fig:002_module_faces}
	\end{subfigure}
    \begin{subfigure}[t]{\textwidth}
        \centering
        \includegraphics[width=0.4\linewidth]{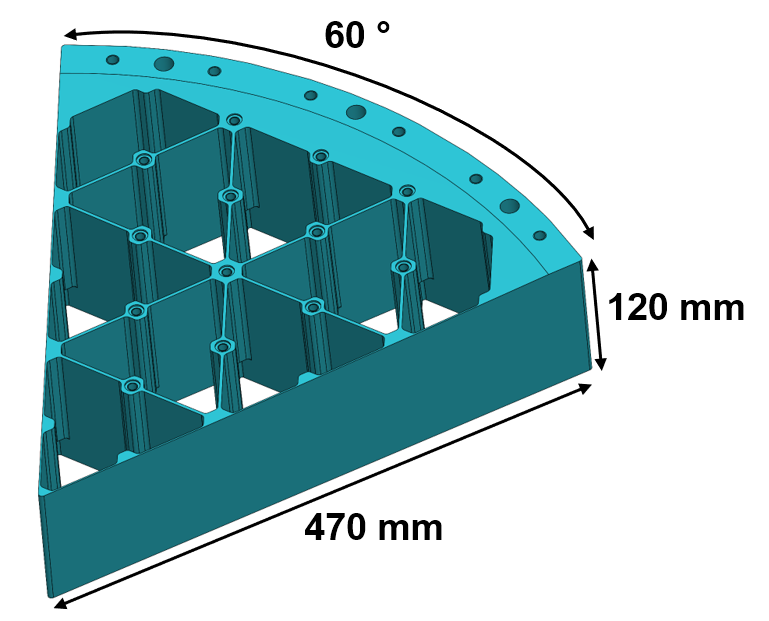}
        \caption{Dimensions overview of the focal plate prototype}
        \label{fig:002_focal_plate cad}
	\end{subfigure}
\vspace{0.2cm}
 \caption{Overview of the plate main design principles}
 \label{fig:002_focal_plate_design principles}
\end{figure}

\noindent The focal plate structure aims for 2 main design principles:
\begin{itemize}
    \item \textit{Semi-frameless design}: as seen in Figure \ref{fig:002_FP_layout_60deg} we partially remove intermediate walls between modules, packing them closely by intermediate groups of 4, and thus optimizing the packing density of a focal plane layout. While this comes at a cost of a stiffness loss for the structure, we wanted to assess the manufacturability of this design since it can improve the vignetting coverage up to 10\% depending on the project parameters considered.
    \item \textit{Three screws interface}: the modules are inserted from the bottom of the plate. They interface with the structure with three support faces that provided M2.5 threaded holes to be bolted on. This architectrure is kept from the original proposition in Silber, et al (2026)\cite{silber_25000_2022} and reminded in Figure \ref{fig:002_module_faces}
\end{itemize}

\subsection{Manufacturing challenges and process}

Past focal plate structures have been manufactured using 5-axes CNC machining and proved to perform according to specifications for individual robots-based instruments such as SDSS-V \cite{pogge_robotic_2020}, DESI\cite{duan_focal_2019} or MOONS\cite{gonzalez_moons_2022}. Since the problem to be solved is similar for the future module-based instruments, we chose to machine the prototype out of a single block of material. In addition, a large DMG DMF 180 (5 axes) is available in the EPFL Institute of Physics workshop, thus easing in-house manufacturing.\\
The material was chosen to be Aluminium 7022 T651 for its excellent manufacturing stability. The milling process is going to remove almost 90\% of the raw material block leaving relatively thin walls ($\approx$ 2.5 mm) compared to the size of the part, see Figure \ref{fig:002_plat_ass_overview}. Its yield strength of about 390MPa and Brinell hardness of 130 will also be an asset to limit local permanent deformations upon module fastening.\\

\begin{figure}[H]
    \centering
    \includegraphics[width=0.7\linewidth]{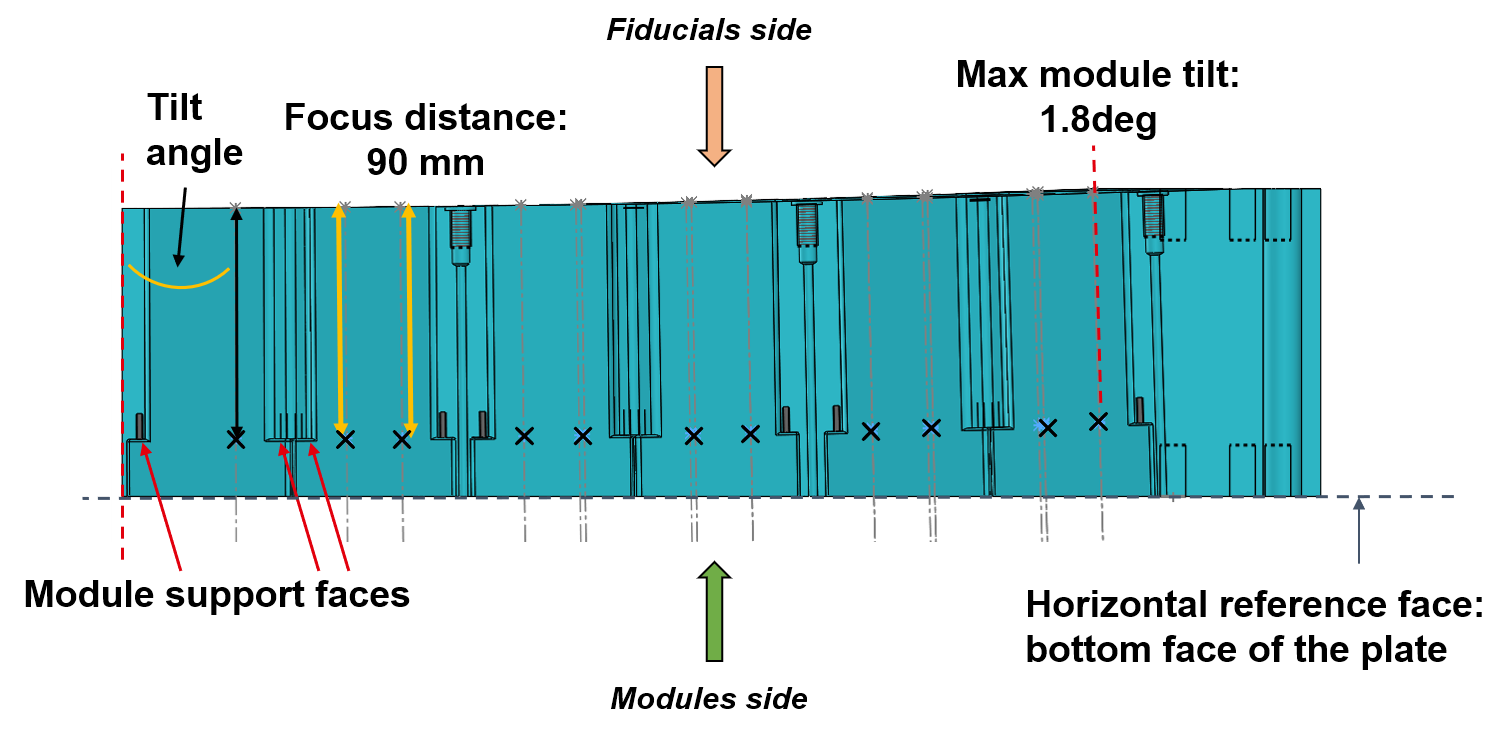}
    \caption{Cut view of the plate in the middle}
    \label{fig:002_cut_view}
\end{figure}

\noindent The chosen plate design, manufacturing technique and material lead to several manufacturing challenges:

\begin{enumerate}
    \item \textbf{Tilt}: modules and fiducials support faces need to be angled by typical increments of 0.2 deg with respect to the vertical axis. Consequently, each triangular pocket needs to be milled to the tilt angle of the corresponding module. Those angles can be seen in Figure \ref{fig:002_cut_view}.
    \item \textbf{Focus}: modules support faces are set to be at a nominal distance of 90 mm from the curved top surface, as pictured in Figure \ref{fig:002_cut_view}. The fiducials support faces are set to be consistently 1.5 mm from the same surface.
    \item \textbf{Focus and tilt tolerances}: at the time of the design of this plate, none of the projects mentioned in theintroduction had settled its specifications. However, typical tolerances, in Table \ref{tab:Manufacturing tolerances}, for such a part were deemed to be a good first approach
    \begin{table}[H]
        \centering
        \caption{Plate manufacturing tolerances}
        \label{tab:Manufacturing tolerances}
            \begin{tabular}{llc} 
            \toprule
             & Value & Unit \\
            Tilt & $\pm$0.05 & ° \\
            Focus & $\pm$30 & $\mu$m \\
            \bottomrule
            \end{tabular}
    \end{table}

    \item \textbf{Walls size}: the separation walls are 2.5 mm thick, see Figure \ref{fig:002_plat_ass_overview} which may induce vibrations during milling. The tilt mentioned in the first point also implies that the walls are slightly tapered: 2.5 mm thick on the top surface and 3 mm on the bottom surface
    \item \textbf{Plate thickness}: the plate is 120 mm thick, thus needs several part flips during the manufacturing process to keep the tools length as short as possible
    \item \textbf{Curved top surface}: although not used for positioning the top surface of the plate is milled to the radius of curvature mentioned in Table \ref{tab:focal_surf_param}. This allows not only to save mass, but also for the positioners to protrude out of the plate from the same amount. 
\end{enumerate}

\begin{figure}[H]
\captionsetup[subfigure]{justification=centering}
 \begin{subfigure}[t]{0.5\textwidth}
		\centering
		\includegraphics[width=\linewidth]{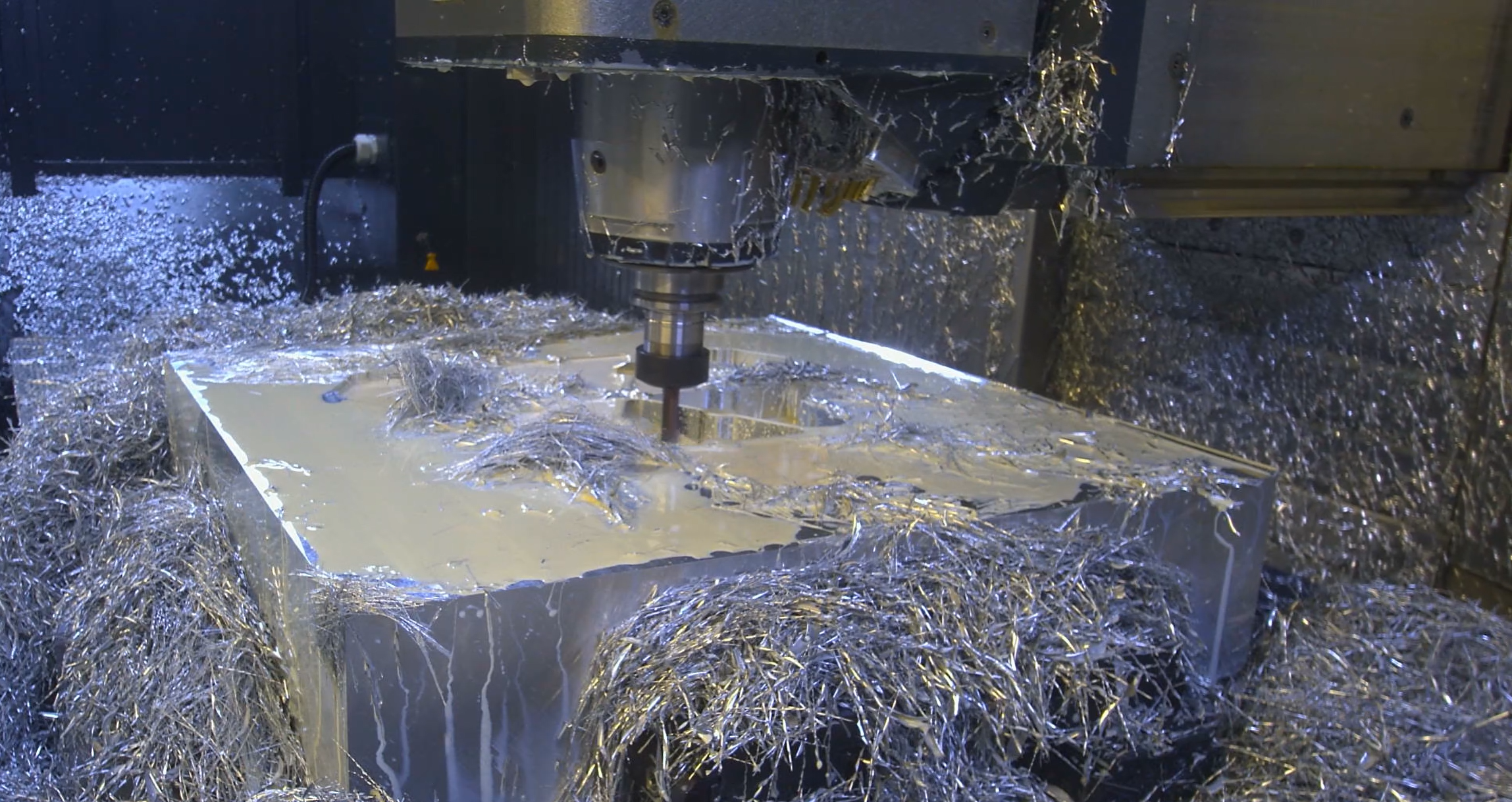}
		\caption{First roughing passes}
        \label{fig:002_rough_pass}
	\end{subfigure}
	\begin{subfigure}[t]{0.5\textwidth}
		\centering
		\includegraphics[width=0.8\linewidth]{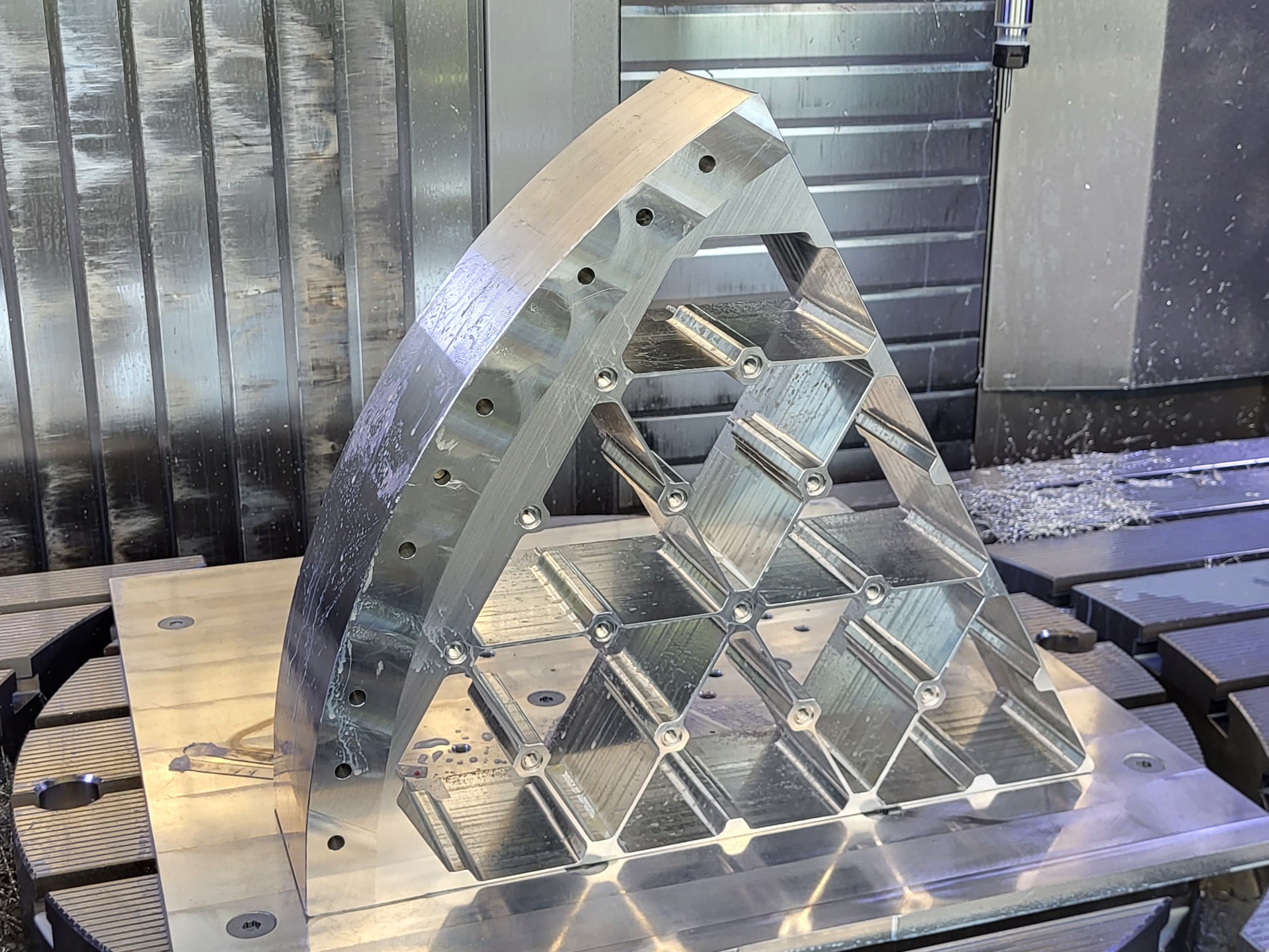}
        \caption{Finalized part facing the top, curved, surface}
        \label{fig:002_final_top}
	\end{subfigure}

        \begin{subfigure}[t]{\textwidth}
        \centering
        \includegraphics[width=0.22\linewidth]{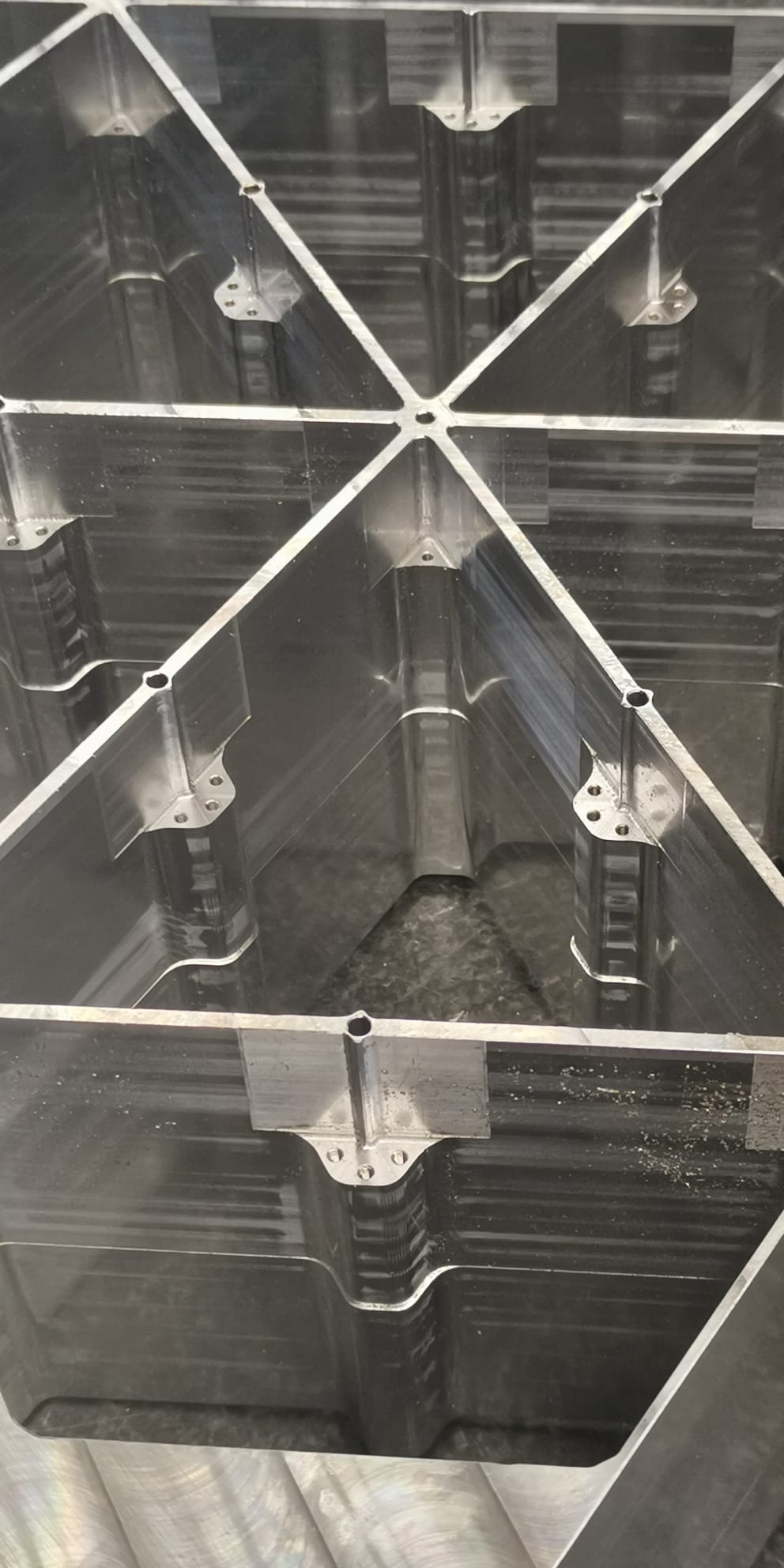}
        \caption{Finalized part facing the bottom view and highlighting the  M2.5-threaded modules support faces and the 3.5mm through holes for fiducials cabling}
        \label{fig:002_final_bottom}
	\end{subfigure}
\vspace{0.2cm}
 \caption{Overview of the focal plate manufacturing process}
 \label{fig:002_focal_plate_final}
\end{figure}

With all those challenges in mind the mechanical workshop of the EPFL Physics Institute manufactured this plate from a raw block of 550x550x130 mm. Figure \ref{fig:002_focal_plate_final} gives an overview of the result of the 5 axes milling process.

\section{CMM Measurements}
\subsection{Part measuring method}
To assess the manufacturing tolerances from Table \ref{tab:Manufacturing tolerances} we chose to measure the plate prototype on a Coordinate Measuring Machine (CMM). The workshop of the EPFL Mechanical Engineering Institute has a dedicated metrology room and owns a COORD3 - HERA 10.07.05. It is sufficiently large to host the plate and has typical precision of $\pm$2.4$\mu m$ with the probe used and the dimensions that we want to measure. Finally, using a CMM allows to easily repeat measurement cycles to build statics and confidence in the reliability of the results, as it will be confirmed in Section \ref{sec:results}.

\begin{figure}[H]
\captionsetup[subfigure]{justification=centering}
 \begin{subfigure}[t]{0.4\textwidth}
		\centering
	\includegraphics[width=0.7\linewidth]{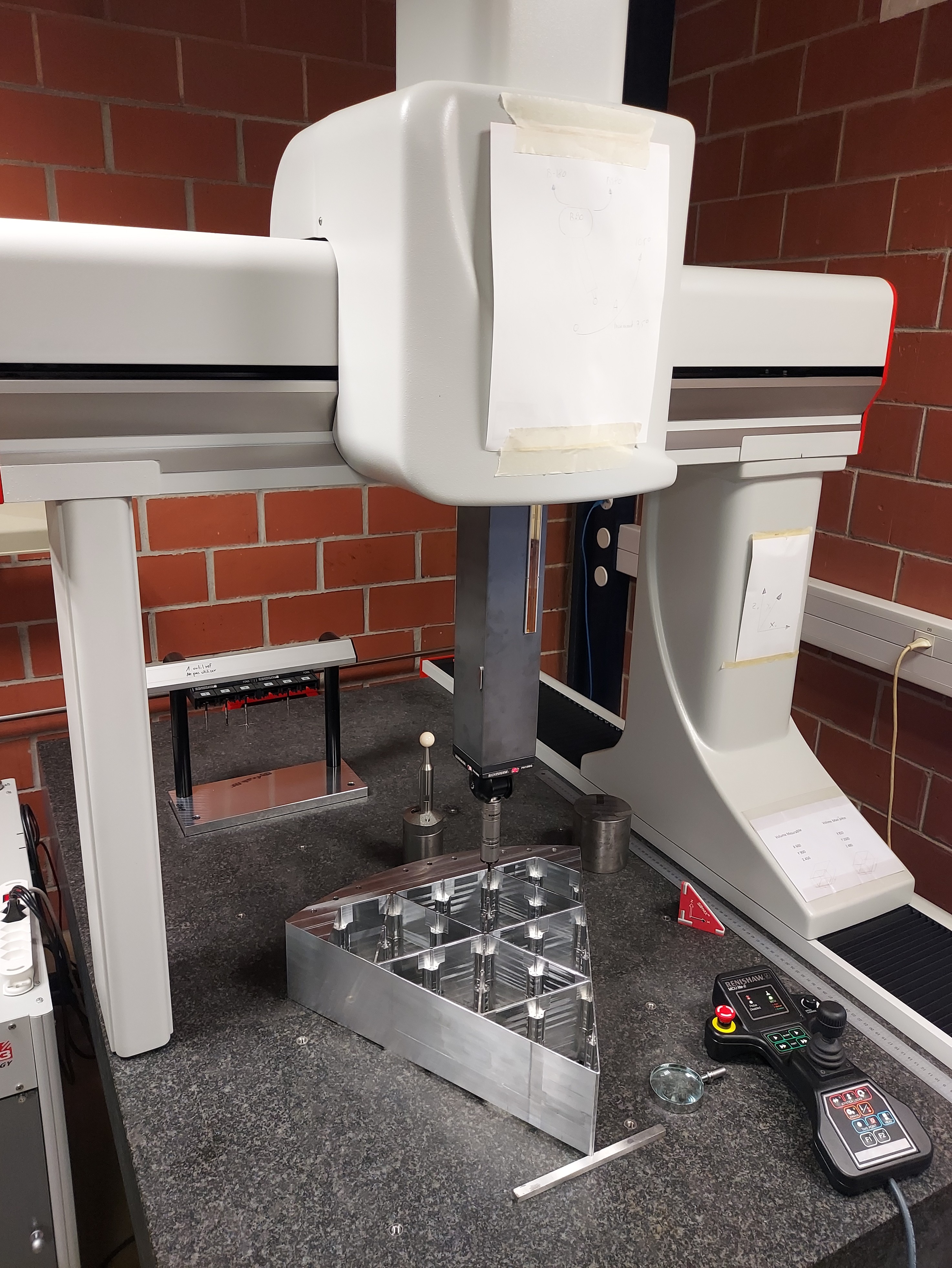}
		\caption{Plate under the CMM}
        \label{fig:003_plate_in_CMM}
	\end{subfigure}
	\begin{subfigure}[t]{0.58\textwidth}
		\centering
		\includegraphics[width=1\linewidth]{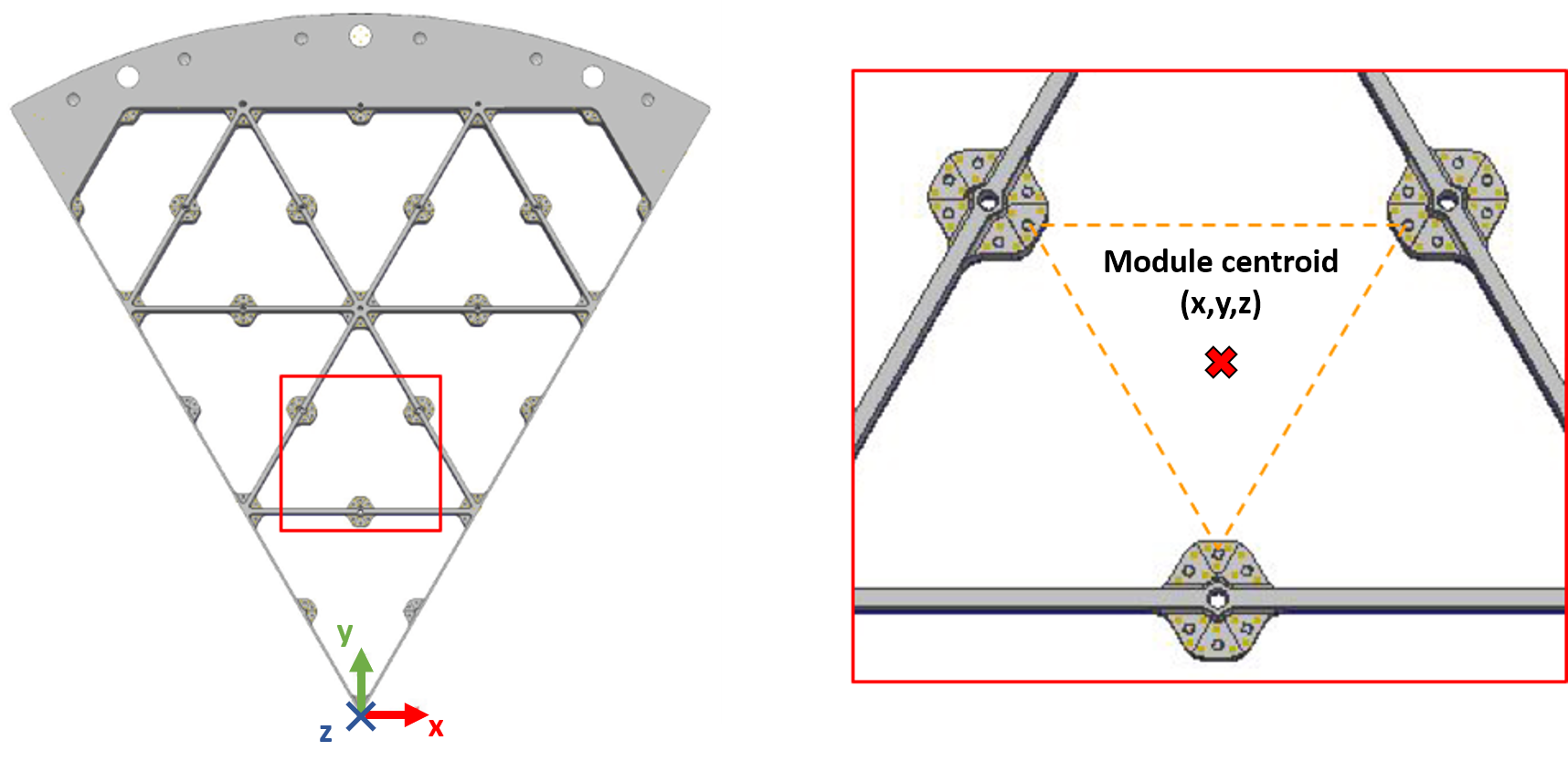}
        \caption{Core measurement principle: probe 3 points per module support face (points in yellow)}
        \label{fig:003_meas_methodo}
	\end{subfigure}
\vspace{0.2cm}
 \caption{Overview of the focal plate measuring process}
 \label{fig:003_plate_measurement}
\end{figure}
\label{sec:meas_methodo}
We started by measuring the \textit{module side} of the plate, with the bottom face oriented upward. The part coordinate system is defined by probing the two lateral 60 degrees faces which defines the z-axis portrayed in Figure \ref{fig:003_meas_methodo}. Probing the flat bottom faces gives us the horizontal plane reference needed to intersect with the z-axis and create the origin of measurements at the tip of the plate as in Figure \ref{fig:003_meas_methodo}.\\
After coordinate system definition, each module support face is probed on 3 points, see the yellow dots in Figure \ref{fig:003_meas_methodo}, from which a local plane is built. The three planes corresponding to one module space are then averaged to obtain an average plane, portrayed in yellow dashed lines in Figure \ref{fig:003_meas_methodo}.\\
A similar approach is done when flipping the part to measure the \textit{fiducials side}, with the bottom face oriented downward. The z-axis is defined in an identical fashion and the horizontal reference is the flat face hosting the fixation features portrayed in Figure \ref{fig:002_plat_ass_overview}.
Each measurement cycle for both side is repeated 5 times to ensure the measurements reliability.

\section{Results - Performances of the manufacturing process}
\label{sec:results}
\subsection{Modules side}
The modules are numbered in ascending order of tilt angle which also correlates with their radial distance to the plate center.
\begin{figure}[H]
    \centering
    \includegraphics[width=0.35\linewidth]{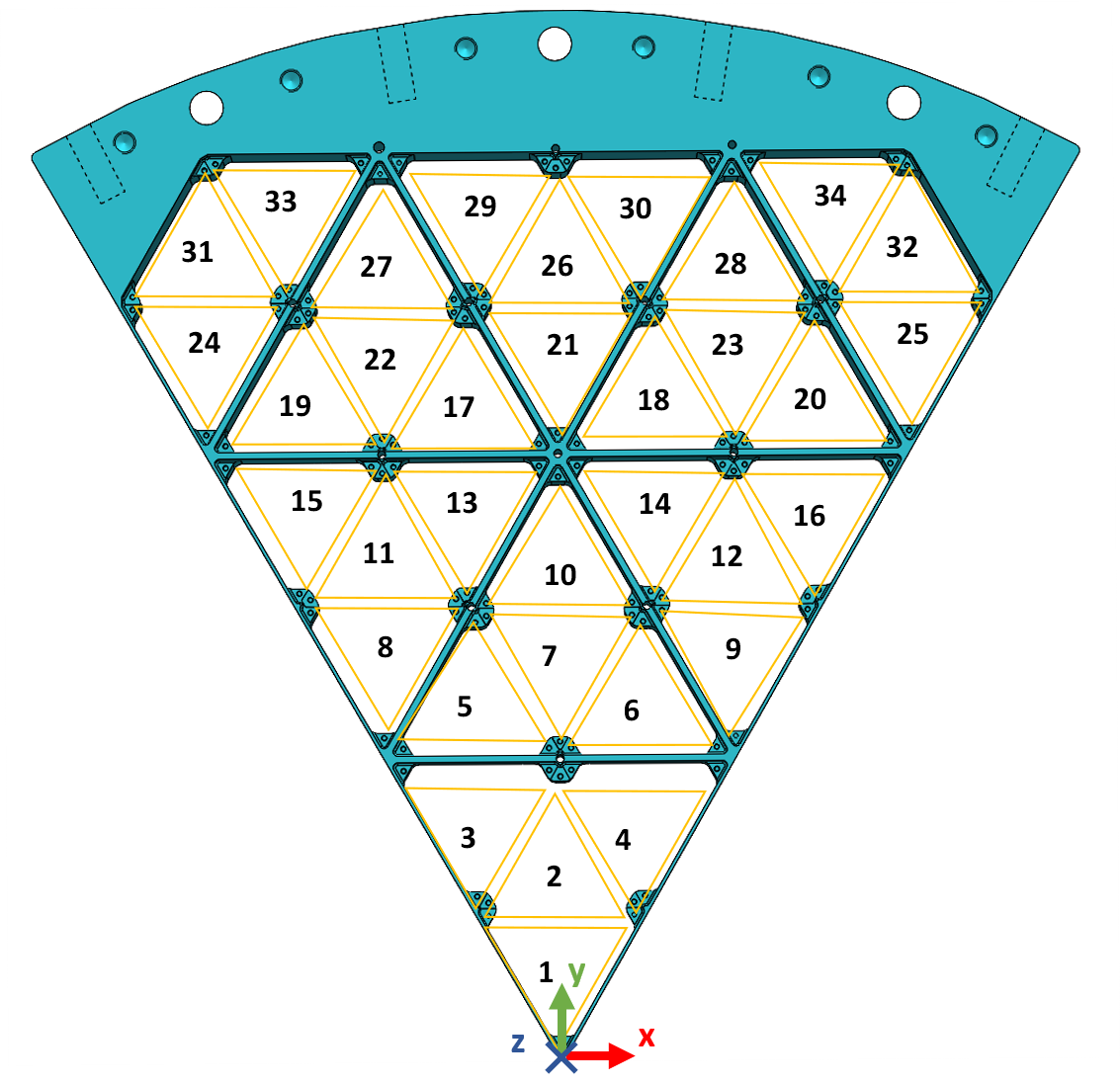}
    \caption{Module numbering from 1 to 34}
    \label{fig:004_modules_number}
\end{figure}

\subsubsection{Module Tilt}
The module tilt is defined as the 3D angle between the average module plane built from the support faces measurements in Section \ref{sec:meas_methodo} and the vertical z axis; also defined as the \textit{polar angle} in spherical coordinates.
\begin{figure}[H]
\captionsetup[subfigure]{justification=centering}
 \begin{subfigure}[t]{0.49\textwidth}
		\centering
	\includegraphics[width=0.87\linewidth]{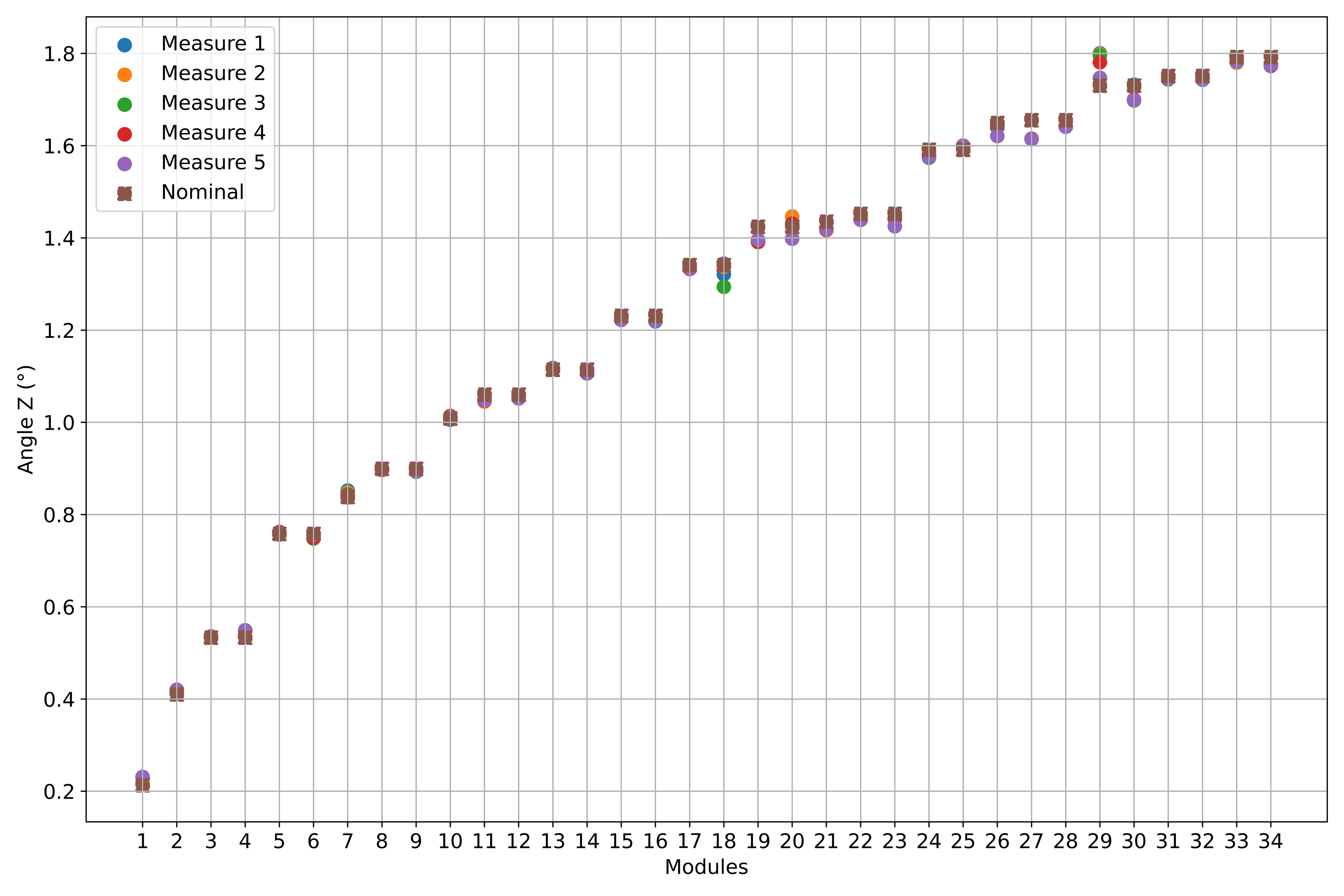}
		\caption{Nominal module angles compared to each measurement cycle }
        \label{fig:004_Angle_nominal_vs_measure}
	\end{subfigure}
	\begin{subfigure}[t]{0.49\textwidth}
		\centering
		\includegraphics[width=\linewidth]{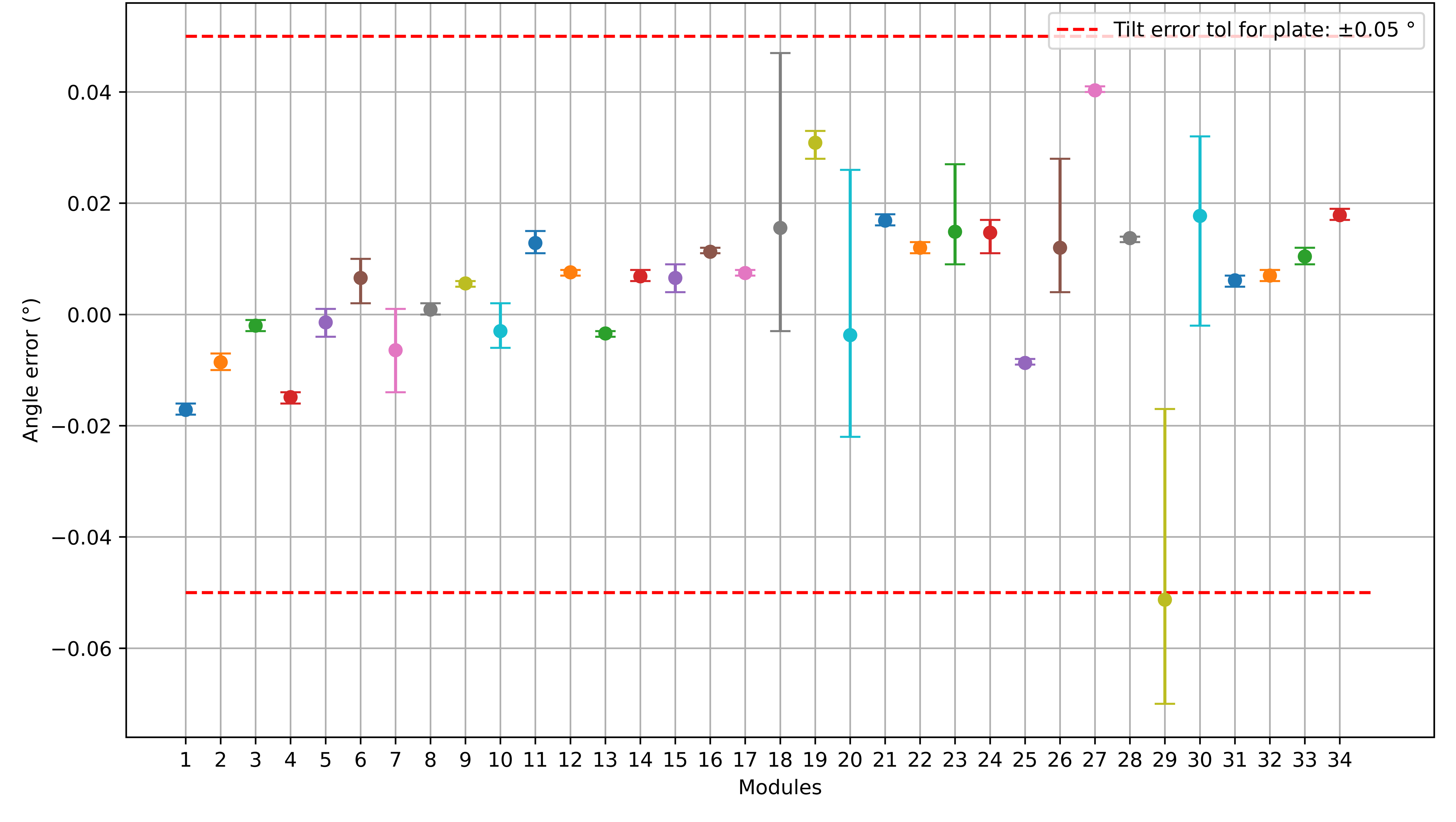}
        \caption{Core measurement principle: probe 3 points per module support face (points in yellow)}
        \label{fig:004_module_tilt_error}
	\end{subfigure}
    \begin{subfigure}[t]{\textwidth}
        \centering
        \includegraphics[width=0.55\linewidth]{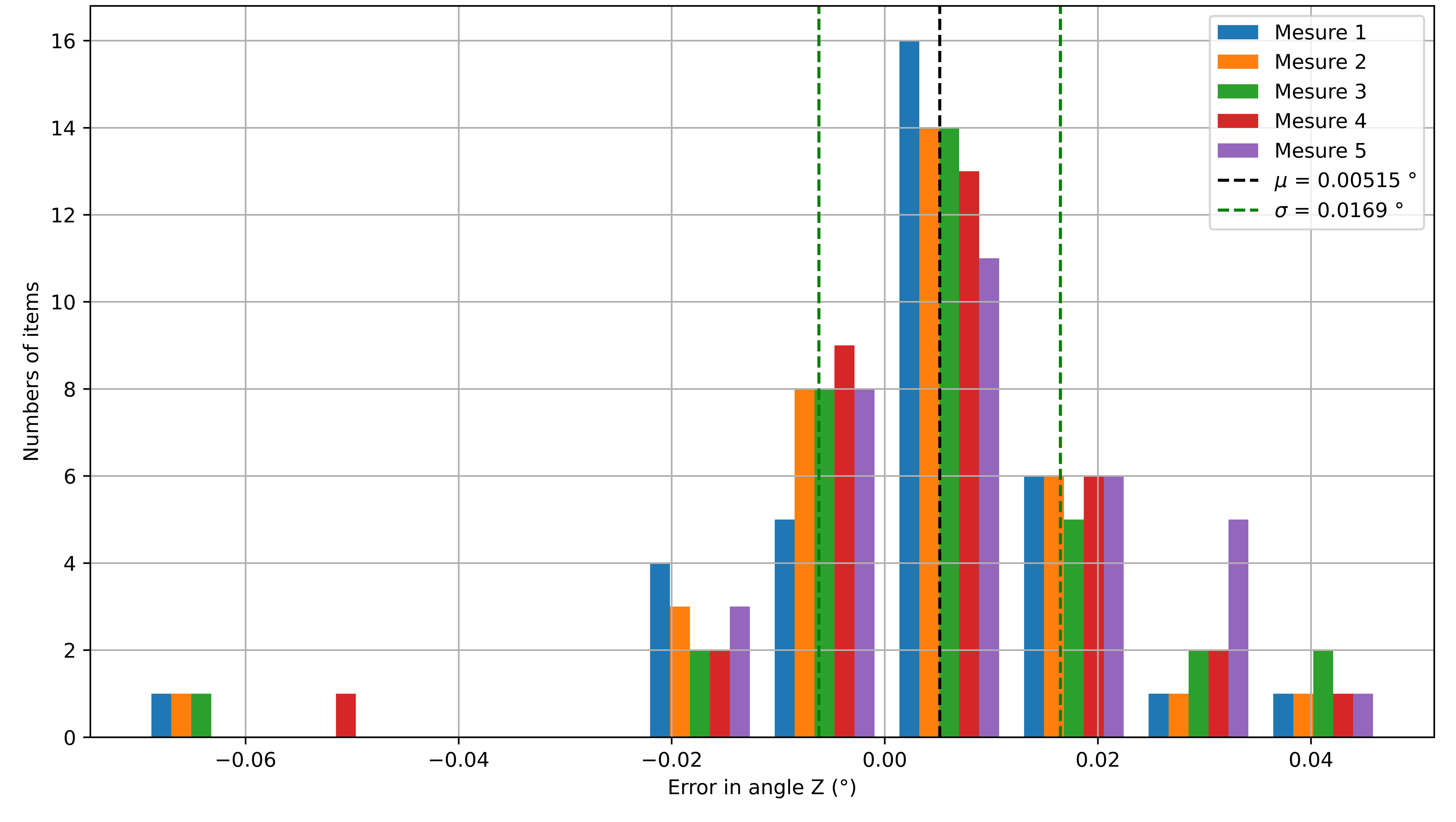}
        \captionsetup{width=0.55\linewidth}
        \caption{Distribution of module focus angle errors; centered on a mean error of $\mu_{tilt} =0.0051$° and a standard deviation of $\sigma_{tilt}=0.017$°}
        \label{fig:004_module_tilt_error_distribution}
	\end{subfigure}
\vspace{0.2cm}
 \caption{Overview of the module support faces nominal angles with respect to z and their manufacturing errors}
 \label{fig:004_module_tilt_results}
\end{figure}

Figure \ref{fig:004_Angle_nominal_vs_measure} shows a first visual incentive how close the measurements, thus the physical part, is from the nominal angles. It highlights that the first module is already tilted by 0.2 deg and the last ones by 1.8 deg.\\
Apart from a few outliers that exhibit a large measurement spread, most angles are confidently confirmed to lie within the tolerance envelope of $\pm$0.05°, see tilt errors plotted in Figure \ref{fig:004_module_tilt_error}. It will be up to the future instrument projects to accept and mitigate those on the full plate or reject the part.\\
Figure \ref{fig:004_module_tilt_error_distribution} gives a 1$\sigma$representation of the spread of the error. It highlights that the tilt error is centered around a mean of $\mu_{tilt} =0.0051$°, indicating almost no bias, and a standard deviation of $\sigma_{tilt}=0.017$°. The latter is often interpreted in 1$\sigma$ representation as 68\% of the considered items are contained within $\mu_{tilt}\pm \sigma_{tilt}$.

\subsubsection{Module Focus}
The focus distance from the module support faces to the top curved surface is set to be 90 mm. Since we know the x,y,z position of the projection of the module centroid on the spherical top surface, we can calculate its 3D Eucildean distance from the module centroid determined by the measurements; see Figure \ref{fig:003_meas_methodo}.
\begin{figure}[H]
\captionsetup[subfigure]{justification=centering}
 \begin{subfigure}[t]{0.49\textwidth}
		\centering
	\includegraphics[width=\linewidth]{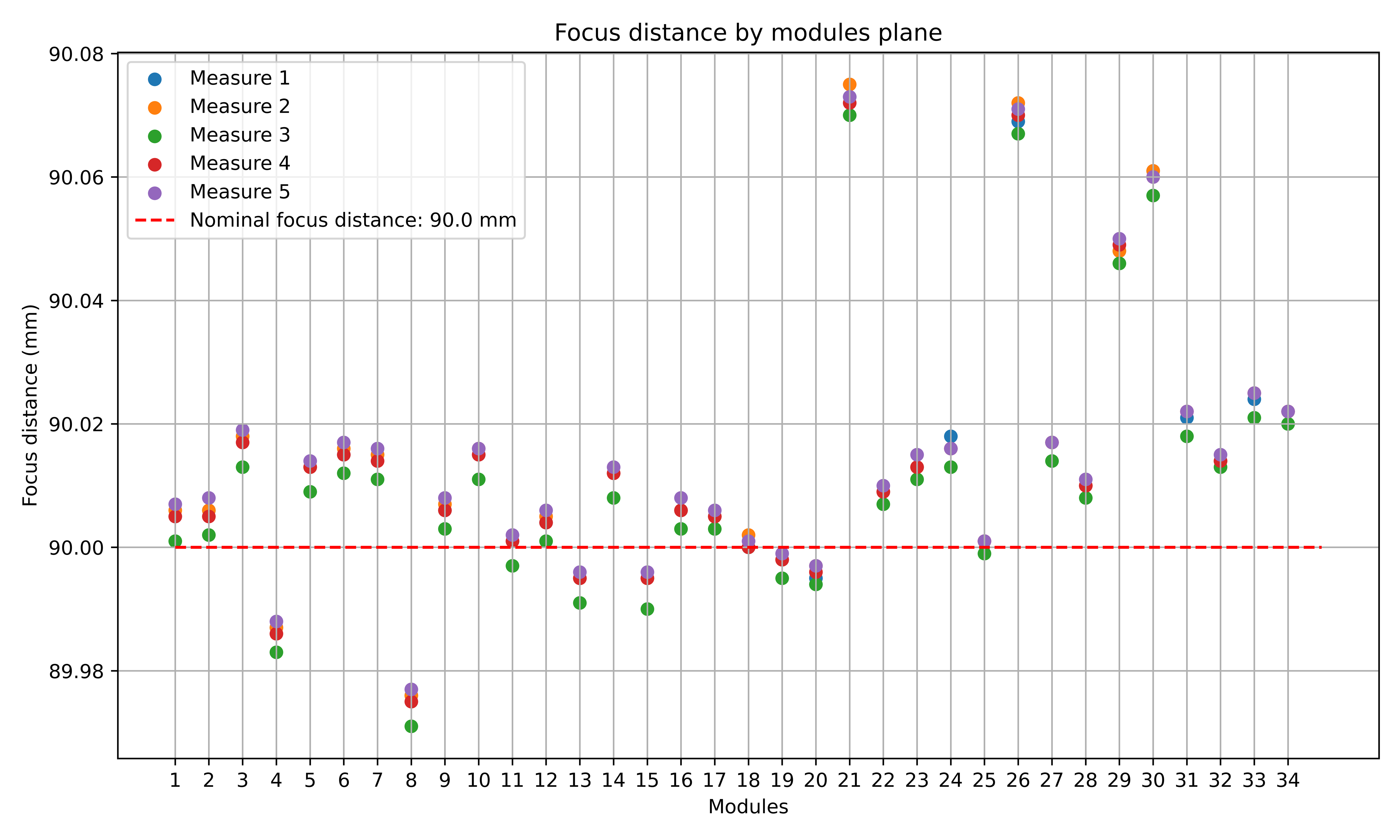}
		\caption{Nominal module focus compared to each measurement cycle }
        \label{004_Focus_nominal_vs_measure}
	\end{subfigure}
	\begin{subfigure}[t]{0.49\textwidth}
		\centering
		\includegraphics[width=\linewidth]{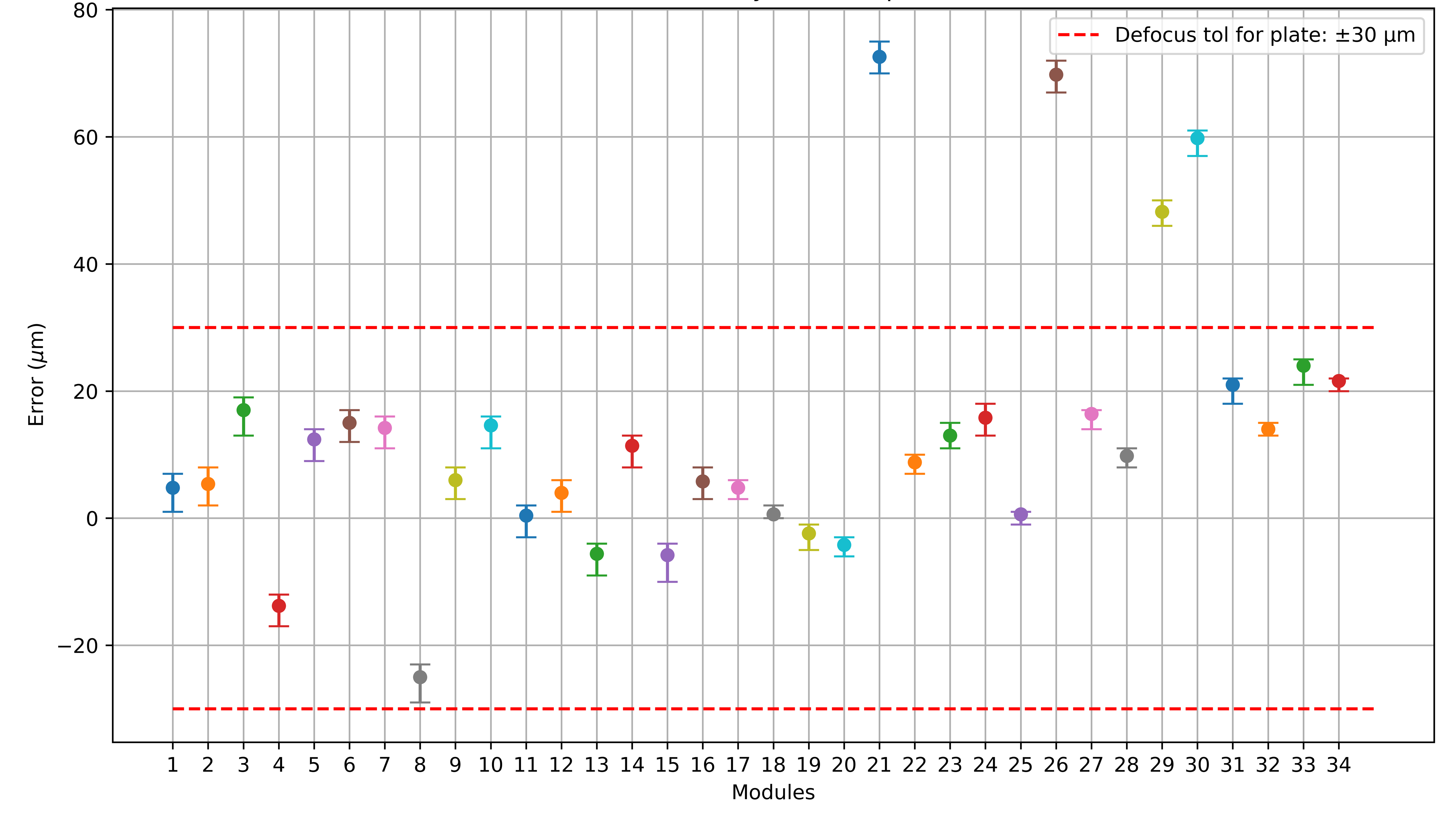}
        \caption{Focus errors for each module support face compared to the desired $\pm$30$\mu m$ tolerance envelope}
        \label{fig:004_module_focus_error}
	\end{subfigure}
    \begin{subfigure}[t]{\textwidth}
        \centering
        \includegraphics[width=0.55\linewidth]{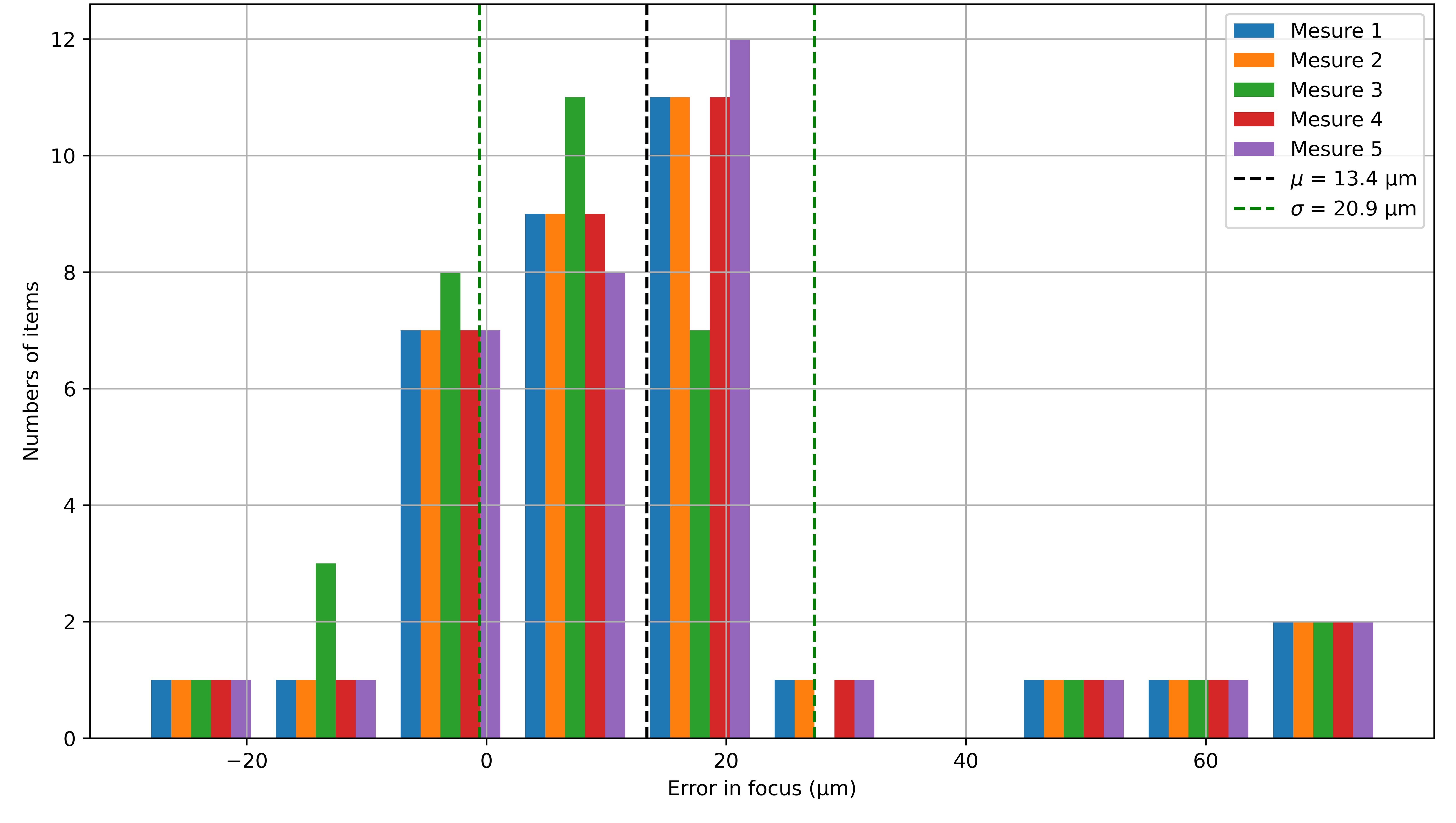}
        \captionsetup{width=0.55\linewidth}
        \caption{Distribution of module focus angle errors; centered on a mean error of $\mu_{focus} =13.4 \mu m$ and a standard deviation of $\sigma_{focus}=20.9\mu m$}
        \label{fig:004_module_focus_error_distribution}
	\end{subfigure}
\vspace{0.2cm}
 \caption{Overview of the module support faces nominal angles with respect to z and their manufacturing errors}
 \label{fig:004_modules_focus_results}
\end{figure}

In a similar fashion as the module tilt representation, Figure \ref{004_Focus_nominal_vs_measure} shows a first order, visual representation of the nominal, expected, focus value and the actual measurement points. They also appear close to nominal, and Figure \ref{fig:004_module_focus_error} confirms that incentive. Here the measurement spread for each module is a few microns. We can see a few outliers again for module 21, 26, 29 and 30, compared to the focus tolerance envelope of $\pm 30 \mu m$. Those can me mitigated via the shim stacks on the three module vertices observed in Figure \ref{fig:002_module_faces}.\\
The distribution shown in Figure \ref{fig:004_module_focus_error_distribution} highlights a small bias of the focus errors of $\mu_{focus} =13.4 \mu m$. This 1$\sigma$ representation also shows us that 68\% of the focus errors are laying within an envelope of $\pm \sigma_{focus}=20.9\mu m$.

\subsection{Fiducials side}
The fiducials are numbered in ascending order of tilt angle which also correlates with their radial distance to the plate center.
\begin{figure}[H]
    \centering
    \includegraphics[width=0.35\linewidth]{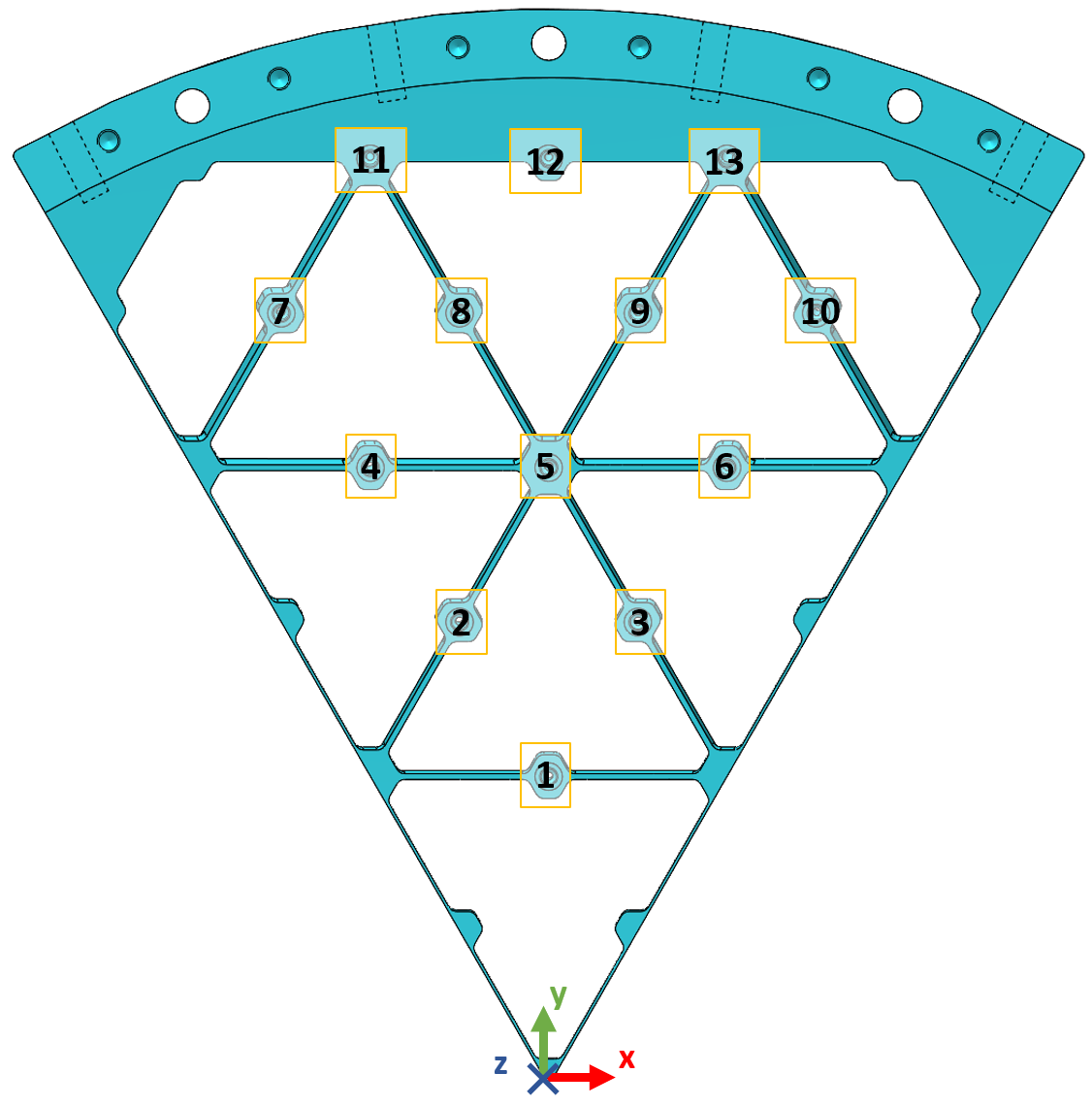}
    \caption{Fiducials numbering from 1 to 13}
    \label{fig:004_fiducials_number}
\end{figure}

The fiducials support faces tilt and focus errors are calculated in an identical manner as previously for the modules.
\begin{figure}[H]
\captionsetup[subfigure]{justification=centering}
 \begin{subfigure}[t]{0.49\textwidth}
		\centering
	\includegraphics[width=\linewidth]{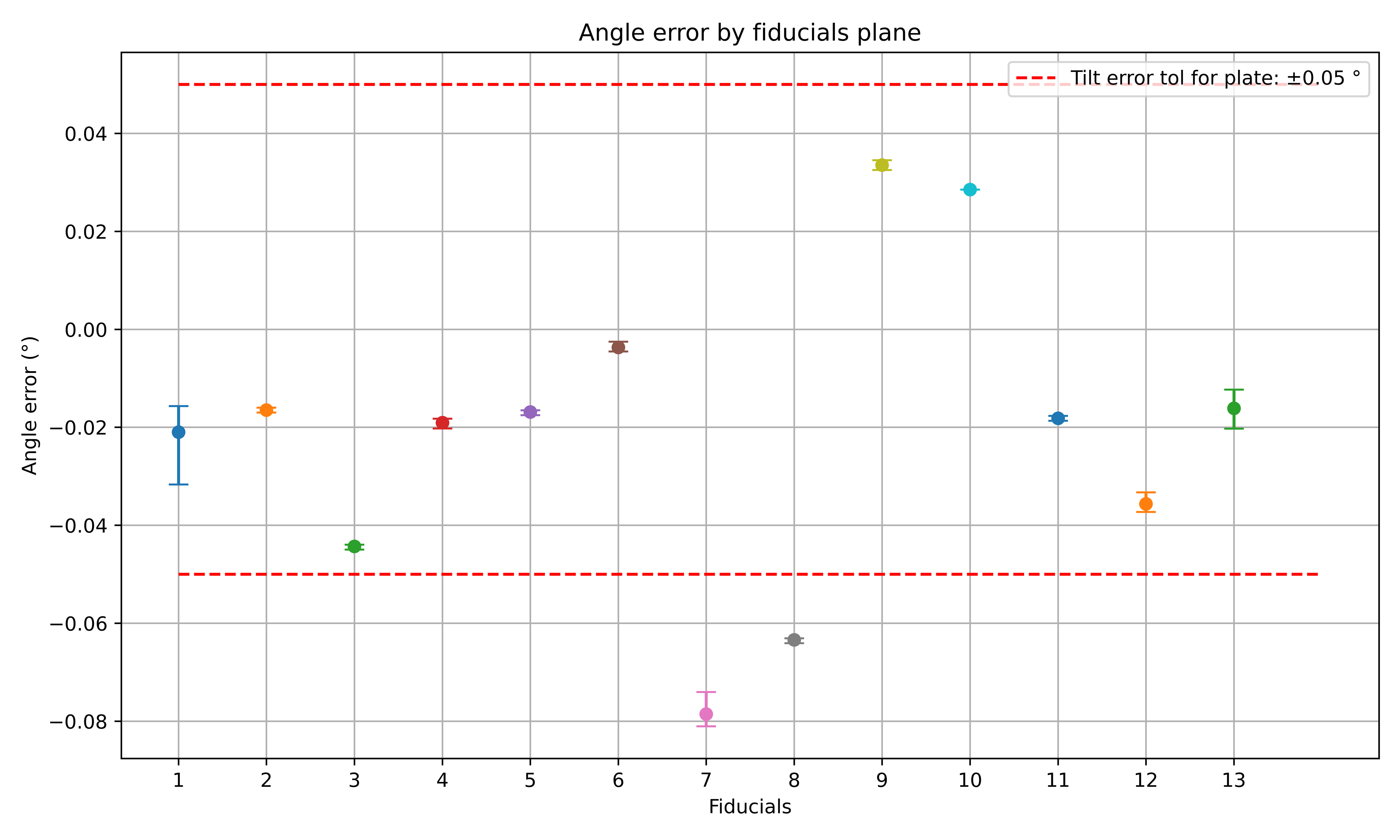}
		\caption{Fiducials support faces angular error}
        \label{fig:004_fiducials_angle_error}
	\end{subfigure}
	\begin{subfigure}[t]{0.49\textwidth}
		\centering
		\includegraphics[width=\linewidth]{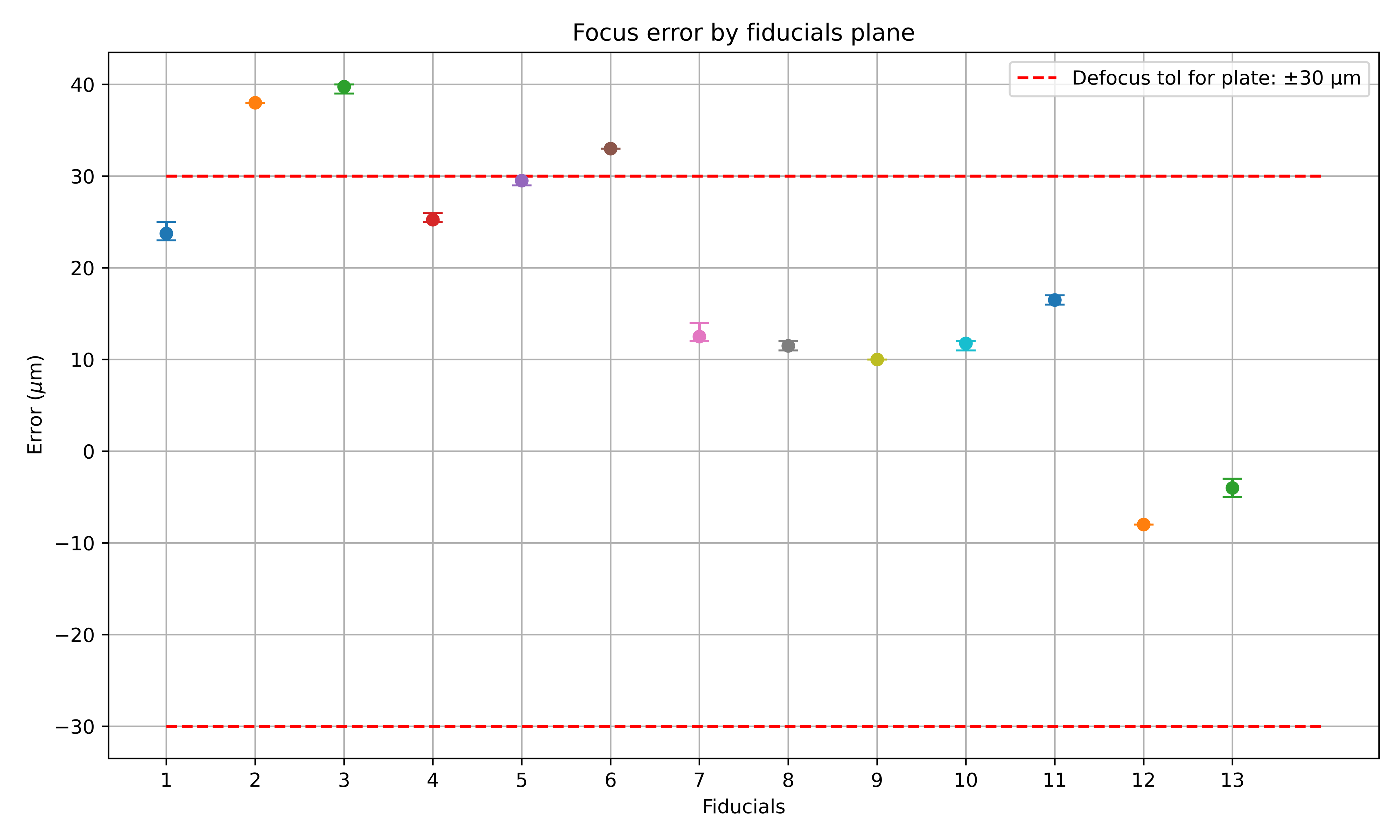}
        \caption{Fiducials support faces focus error}
        \label{fig:004_fiducials_focus_error}
	\end{subfigure}
\vspace{0.2cm}
 \caption{Overview of the fiducials support faces tilt and focus errors alongside tolerance envelopes}
 \label{fig:004_fiducials_results}
\end{figure}

Apart from a few outliers the fiducial support faces also exhibit good performances compared to the manufacturing tolerance envelope with enough confidence in the measurements, as seen in Figure \ref{fig:004_fiducials_results}.
\section{Conclusion}
In conclusion, the presented work shares one possible triangular module layout design for future Stage-5 instruments. A few simplifications were made such as assuming no CRD or a spherical focal surface. Those are the consequences of the current study not being specifically realized in the scope of one of the Stage-5 project mentioned in the introduction but in the broader R\&D framework of \textit{Innosuisse} in Switzerland. Thus, it does not pretend to impose any of the methods presented here, just sharing thoughts and results.\\
The modeling, manufacturing and measurement methods of the focal plate prototype used here proved to be a promizing solution for future triangular module-based focal planes. The results are well within the tolerance envelopes.\\
Future work will be conducted to test module integration repeatability and assess how reliably this precisely manufactured plate can maintain those future fiber positioners.

\acknowledgments 
The authors acknowledge Jean-David Perriard and his team from the EPFL IPHYS mechanical workshop for the astonishing manufacturing work to build that focal plate. They showcased their expertise and proved that it is possible to build such a complex part.\\
We also aknowledge Laurent Chevalley from the EPFL Mechanical Engineering Institute workshop for his patience and careful work on the CMM measurement of this focal plate prototype.\\
Finally this work was originally set in motion by the support of Innosuisse - Swiss Innovation Agency under Grant Agreement No. 101.014 IP-ENG.
 

\bibliography{references_FIXED}

\bibliographystyle{spiebib} 


\end{document}